\documentclass[twocolumn,prl,showpacs,floatfix]{revtex4}

\usepackage{graphicx}
\usepackage{dcolumn}
\usepackage{bm}
\usepackage{amsmath}

\begin{document}

\title{Nonuniversal transmission phase lapses through a quantum dot:\\
An exact-diagonalization of the many-body transport problem
}

\author{Leslie O. Baksmaty}
\author{Constantine Yannouleas}
\author{Uzi Landman}

\affiliation{School of Physics, Georgia Institute of Technology,
             Atlanta, Georgia 30332-0430}

\date{13 December 2008; Phys. Rev. Lett. {\bf 101}, 136803 (2008)}

\begin{abstract}
Systematic trends of nonuniversal behavior of electron transmission phases 
through a quantum dot, with no phase lapse for the transition  
$N=1 \rightarrow N=2$ and a lapse of $\pi$ for
the $N=2 \rightarrow N=3$ transition, are predicted, in agreement with
experiments, from many-body transport calculations involving exact
diagonalization of the dot Hamiltonian. The results favor anisotropy of the 
shape of the dot and strong $e-e$ repulsion with consequent electron
localization, showing dependence on spin configurations and the 
participation of excited doorway transmission channels.
\end{abstract}

\pacs{73.23.Hk, 73.21.La, 31.15.-p}

\maketitle

{\it Motivation.\/} 
Measurements via Aharonov-Bohm interferometry of transmission phases (in addition
to conductances) provide valuable information about fundamental transport 
properties through small systems \cite{heib97,heib05}. 
An experimental setup conceived and employed in 
such studies consists of placing a two-dimensional quantum dot (QD) in one 
of the arms of the two-path interferometer \cite{heib97,heib05}. 
In earlier experiments \cite{heib97} a universal behavior 
of the phase was found where for each added electron the phase drops 
discontinuously by $\pi$ (phase lapse, PL) before rising back (as expected) 
continuously by the same amount. This behavior was found to be independent of 
the number of electrons, the dot shape and spin degeneracy. The regime of 
mesoscopic (nonuniversal) behavior  of the transport through the dot, exhibiting
an irregular PL pattern and dependencies on the number of electrons and dot 
parameters has been observed only recently \cite{heib05} for dots with a 
smaller number of electrons $N < 15$.
The large majority of theoretical studies attempted to address the PL 
behavior in the universal regime. Nevertheless, this behavior remains puzzling 
and challenging \cite{njp07,karr07}. 

Here, we focus on the nonuniversal regime (small $N$) where ``the phase behavior 
for electron transmission should in principle be easier to interpret''
\cite{heib05}. To this aim, we use a transport theory, based on a computational 
approach entailing exact diagonalization (EXD) \cite{rop07}
of the QD many-body Hamiltonian. This approach includes electron correlation 
effects and allows systematic investigations of the transport 
characteristics as a function of the dot shape, number of electrons, and spin 
configurations. Specifically, we follow the work of 
Bardeen \cite{bard61} where the correlated many-body states of the QD play a 
determining role (see below) through the socalled quasi-particle wave functions
\cite{bard61,gurv} [see Eq.\ (\ref{wfqp})], replacing the single-particle
orbitals used in independent-particle (and/or mean-field) approaches 
\cite{weid96,hack01}. This
allows evaluation of both the current and transmission phase lapse. Here
our focus is on the phase-lapse phenomenon \cite{note0}.

The transmission probability (current) is given \cite{bard61} (to lowest order) 
by the golden rule as a product of the square
of a dot-to-lead coupling matrix element, ${\cal M}$, and a density of states
factor \cite{kina92}. Exploration of the PL phenomenon in electron
transmission from a right (R) to a left (L) lead through a weakly coupled dot
involves (as part of ${\cal M}$) the product of correlated quasi-particle
wave functions of the dot evaluated inside the R and L lead-to-dot barriers.
${\cal M}$ also involves products of the tails of the R and L lead states
in the barriers, but these depend only weakly (for high barriers) 
on the experimentally used plunger potential, and thus they do not
enter PL considerations \cite{heib05,bard61,gurv,weid96}, when $N$ changes.

In the nonuniversal regime, it is desirable to have precise knowledge about 
the dot parameters (e.g., shape anisotropy and strength of interelectron 
repulsion). While such characterization has not been done in Ref.\ \cite{heib05},
we inquire here about trends that appear in the calculations as a function of the
dot parameters, with subsequent comparisons with the available data. The two key 
parameters that we vary are: (a) the dielectric constant $\kappa$; 
driving the QD towards the regime of strong correlations, since the 
QDs in Ref.\ \cite{heib05} are shallower than usual (i.e.,
with larger $R_W$, see below), and (b) the 
anisotropy of the QD, since the sideway plunger influences greatly the 
average confinement for small dots \cite{elle06}.

We performed a systematic examination of  the 
$N = 1 \rightarrow N = 2$ and $N = 2 \rightarrow N = 3$ transitions. In addition 
to being computationally least prohibitive in the EXD method, these transitions
exhibit the principal generic features
observed in the nonuniversal regime. In particular, for the first transition 
no PL was measured, while a phase lapse was found 
for the second one. Our calculations reproduce these findings
for an appropriate range of dot parameters. We
predict that: (1) Agreement with the experiment is obtained when the dot 
parameters (shape anisotropy and $e-e$ interaction strength) are chosen to be 
favorable for electron localization (formation of electron molecules).
(2) Spin states are important in the selection of transport channels. (3) Excited 
doorway states play a key role.

{\it Theory.\/}
The current intensity and the electron-transmission phase through the quantum dot
relate to a quasiparticle-type wave function that can be extracted from
the many-body EXD states as \cite{bard61,gurv}
\begin{equation}
\varphi_{\text{QP}} ({\bf r}) = \langle \Phi^{\text{EXD}}_{N-1} |
\psi({\bf r}; \sigma) | \Phi^{\text{EXD}}_{N} \rangle,
\label{wfqp}
\end{equation}
where the single-particle operator $\psi({\bf r}; \sigma)$ annihilates an 
electron with spin-projection $\sigma$ at position ${\bf r}$. For calculating the
quasiparticle orbital $\varphi_{\text{QP}} ({\bf r})$, one uses
\begin{equation}
\psi({\bf r}; \sigma)=\sum_{i=1}^K \phi_i({\bf r}) a_i(\sigma),
\end{equation}  
where $a_i(\sigma)$ are the annihilation operators in the Fock space and 
$\phi_i({\bf r})$ are the single-particle eigenstates of the two-dimensional
anisotropic-oscillator potential 
$V(x,y)=m^*(\omega_x^2 x^2+\omega_y^2 y^2)/2$ that confines the electrons
in the quantum dot, with $K=54$ states in the single-particle basis 
(which guarantees numerical convergence, see Ref.\ \cite{li07}). 

The EXD many-body wave functions are expressed as a linear superposition over
Slater determinants $D$'s constructed with the spin orbitals 
$\phi_i({\bf r})\alpha$ and $\phi_i({\bf r})\beta$, with $\alpha$ $(\beta)$ 
denoting up (down) spins, i.e., 
\begin{equation}
\Phi^{\text{EXD}}_{N} (S, S_z; k)= \sum_I C_I^N(S, S_z; k) D^N(I; S_z).
\label{wfexd}
\end{equation}  
While the Slater determinants conserve only the projection
$S_z$ of the total spin, the EXD wave functions in Eq.\ (\ref{wfexd}) are
eigenfunctions of the square 
${\bf S}^2$ of the total spin, since the many-body Hamiltonian ${\cal H}$  
commutes with ${\bf S}^2$; ${\cal H}=
\sum_{i=1}^N [{\bf p}_i^2/(2m^*) + V(x_i,y_i)]+ \sum_{i<j} e^2/(\kappa r_{ij})$.
$k=1,2,3,\ldots$ counts the ground and excited states.

Since in the experiments in the nonuniversal regime the width of the levels of 
the QD is much smaller than the level separation in the dot (weak dot-lead 
coupling; see p. 531, left column in Ref.\ \cite{heib05}), the QD levels do not 
overlap, thus reinforcing our focus on the dependence
of the transmission phase on the quasiparticle states of the QD 
[Eq.\ (\ref{wfqp})]. The current intensity through the quantum dot 
is proportional \cite{bard61,kina92} to the quasiparticle weight  
\begin{equation} 
{\cal W}= \int |\varphi_{\text{QP}} ({\bf r})|^2 d{\bf r}
\label{wqp}
\end{equation}
The transmission phase lapse 
$\theta_{\text{PL}}$ through the quantum dot is given by   
\begin{equation}
\theta_{\text{PL}} = \Delta \theta_{\text{QP}} - \pi,
\label{the}
\end{equation}
where $  \Delta \theta_{\text{QP}} = \theta_{\text{QP}}(x_L,0) -
\theta_{\text{QP}}(x_R,0)$; $\Delta \theta_{\text{QP}}=0$ if 
$\Re[\varphi_{\text{QP}} ({\bf r}_L)*\varphi_{\text{QP}} 
({\bf r}_R)]>0 $ 
and $\Delta \theta_{\text{QP}}=\pi$ if 
$\Re[\varphi_{\text{QP}} ({\bf r}_L)*\varphi_{\text{QP}} 
({\bf r}_R)]<0 $ 
\cite{gurv}, with ${\bf r}_L$ and ${\bf r}_R$ being the positions 
specifying the left and right potential barriers that demarcate the quantum dot.

{\it Results.\/}
We study the evolution of the quasiparticle weight ${\cal W}$ and the 
transmission phase $\theta_{\text{QP}}({\bf r})$ as a function 
of the anisotropy parameter
$\eta=\omega_x/\omega_y$ and the strength of correlations expressed via the
Wigner parameter $R_W$ $(\kappa)$ for the two transitions $N =1 \rightarrow
N=2$ and $N=2 \rightarrow N=3$; 
$R_W (\kappa)= (e^2/\kappa l_0)/\hbar \omega_0$ is the ratio between 
the $e-e$ repulsion and the average energy quantum of the 
confinement [$l_0=\sqrt{\hbar/(m^* \omega_0)}$ is the characteristic length; 
$\omega_0=\sqrt{(\omega_x^2+\omega_y^2)/2}$].

\begin{figure}[t]
\centering{\includegraphics[width=6.50cm]{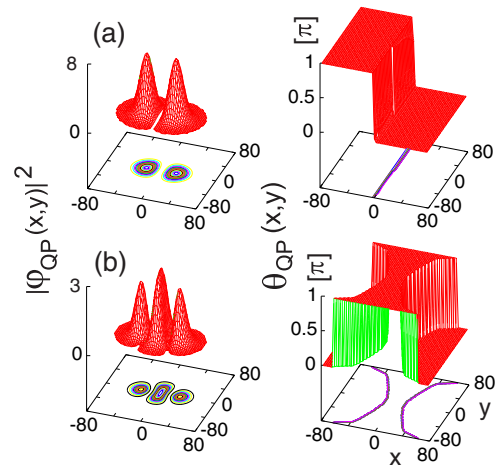}}
\caption{
Quasiparticle wave functions [$|\varphi_{\text{QP}} ({\bf r})|^2$ in units
of $10^{-4}/$nm$^2$, left, and position-dependent phases 
$\theta_{\text{QP}}({\bf r})$, right] for the $N=2 \rightarrow N=3$ transition
in a quantum dot with anisotropy $\eta=0.724$. 
$\theta_{\text{QP}}({\bf r})$ is 0 or $\pi$, since $\varphi_{\text{QP}}$
can be chosen real for zero magnetic field.
The initial state is the ground 
state ($S=0,S_z=0$) for $N=2$ electrons while the final state is (a) the ground 
state ($S=1/2,S_z=1/2$, $E_1=20.888$ meV) and (b) the second excited state 
($S=1/2,S_z=1/2$, $E_3=23.013$ meV) for $N=3$ electrons. 
Lengths in nanometers.
$\hbar \omega_x = 4.23$ meV, $\hbar \omega_y = 5.84$ meV, $\kappa=12.50$ 
(GaAs), and $m^*=0.067$ (GaAs).
}
\label{figqp}
\end{figure}

We display in Fig.\ \ref{figqp} examples of quasiparticle
wave functions (amplitude and phase) for the $N=2 \rightarrow N=3$ transition
for two different final states (corresponding to the 
transitions marked 1 and 3 in Fig.\ \ref{figbars23} below). 
In Fig.\ \ref{figqp}(a), $|\varphi_{\text{QP}}|^2$ exhibits
a two-peak structure separated by one node, while 
$\Delta \theta_{\text{QP}}=\pi$ (with $x_L=-80$ nm and $x_R=80$ nm), 
indicating no phase lapse [see Eq.\ (\ref{the})]. 
In Fig.\ \ref{figqp}(b), $|\varphi_{\text{QP}}|^2$ exhibits
a three-peak structure separated by two nodes, while  
$\Delta \theta_{\text{QP}}=0$, indicating a PL of
$\pi$ [see Eq.\ (\ref{the})]. The nodal structure reflects the 
good parity of $\varphi_{\text{QP}}$, and it emerges from the many-body EXD 
calculation (being {\it a priori\/} unknown).  

\begin{figure}[t]
\centering{\includegraphics[width=6.50cm]{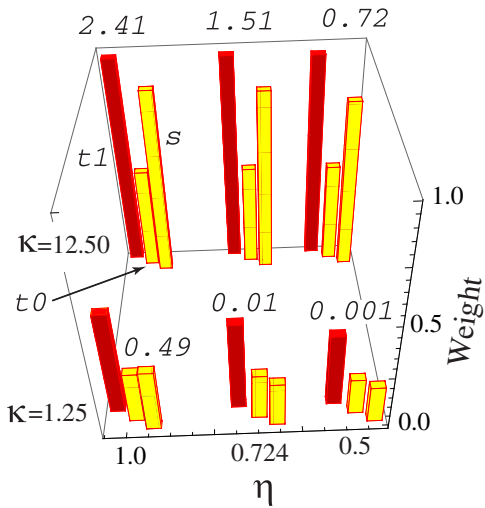}}
\caption{(Color online)
The bar-chart for the quasiparticle of the $N=1 \rightarrow N=2$ transition
in a quantum dot as a function of the strength of the interelectron repulsion
(stronger for smaller $\kappa$) and the anisotropy $\eta$. The 
heights of the bars indicate the weight ${\cal W}$ [Eq.\ (\ref{wqp})], while
shading denotes the quasiparticle transmission
phase [$\Delta \theta_{\text{QP}}$ in Eq.\ (\ref{the})];
dark shade (red) denotes $\Delta \theta_{\text{QP}}=\pi$ (no PL) and gray shade 
(yellow) denotes $\Delta \theta_{\text{QP}}=0$ (occurrence of a PL). 
The bars are arranged in groups of three around each ($\kappa$,$\eta$) point; 
the right bar corresponds to the singlet $(s)$ final state, while the middle and
left ones correspond to the final triplet states with spin projection $S_z=0$ 
($t0$) and $S_z=1$ ($t1$), respectively. The numbers above the bars denote the 
energy difference (in meV) between the ground-state singlet and the excited 
(degenerate) triplets. The effective mass is $m^*=0.067 m_e$ (GaAs). 
For $\eta=1.0$, $\hbar \omega_x=5.0$ meV and $\hbar \omega_y= 5.0$ meV. 
For $\eta=0.724$, $\hbar \omega_x=4.23$ meV and $\hbar \omega_y=5.84 $ meV. 
For $\eta=0.5$, $\hbar \omega_x=3.137$ meV and $\hbar \omega_y= 6.274$ meV. 
Smaller $\kappa$ and $\eta$ enhance electron localization in the dot.
$R_W(\kappa=12.50)=1.51$ and $R_W(\kappa=1.25)=15.1$.}
\label{figbars12}
\end{figure}

We discuss next the $N =1 \rightarrow N=2$ case. The initial state is the ground 
state with a single electron (assuming it has a spin up configuration) occupying
the lowest single-particle level in $V(x,y)$. The final state, 
however, is not restricted to the singlet ground state
[with ($S=0,S_z=0$)] of the $N=2$ quantum dot. Excited states
need to be considered, since an excited ``doorway'' state may be more efficient
(have a higher weight) in transmitting the current through the QD. 
Focussing on the two lowest total energies $E_k(S,S_z)$, we are led to 
consider three final states, i.e., the ground-state singlet 
($S=0,S_z=0$), and the two excited triplets ($S=1,S_z=0$) and ($S=1,S_z=1$),
which are degenerate at zero magnetic field. The third triplet state 
($S=1,S_z=-1$) has zero weight, since flipping of the direction of the initial 
spin is not allowed. The calculated EXD results are displayed in Fig.\ 
\ref{figbars12}. 

The heights of the bars in Fig.\ \ref{figbars12} represent the weight 
${\cal W}$, while the shading (or color online) 
of each bar denotes the quasiparticle transmission phase, with a dark shade 
(online red) denoting $\Delta \theta_{\text{QP}}=\pi$ and a gray shade (online 
yellow) denoting $\Delta \theta_{\text{QP}}=0$. Results are presented for  
a weaker $e-e$ repulsion with $\kappa=12.50$ and a much stronger one with 
$\kappa=1.25$, and for $\eta=1.0$ (circular dot), $\eta=0.724$ (moderate 
anisotropy), and $\eta=0.5$ (strong anisotropy). 

Inspection of Fig.\ \ref{figbars12} reveals the following systematic trends:
(1) The singlet state and the $t0$ triplet have in all instances 
$\Delta \theta_{\text{QP}}=0$ (with a PL $\theta_{\text{PL}}=-\pi$), 
while the $t1$ triplet has $\Delta \theta_{\text{QP}}=\pi$ 
($\theta_{\text{PL}}=0$, i.e., no PL). (2) The excited $t1$ triplet
state has always the largest weight ${\cal W}$, 
and its relative advantage in ${\cal W}$ compared to the $t0$ and the $s$ 
states increases both for stronger correlations (smaller $\kappa$) and stronger 
anisotropies (smaller $\eta$). (3) The singlet-triplet 
energy gap decreases both for smaller $\kappa$ and $\eta$, i.e.,
favoring formation of Wigner molecules.

Consequently, the $t1$ triplet state can act as a doorway state for the
electron transmission in the case of strong correlations and strong anisotropy.
In this case there is no phase lapse ($\theta_{\text{PL}}=0$), 
as observed experimentally \cite{heib05}. Furthermore the experiment
excluded the ground-state singlet as being the final state of the two 
electrons \cite{heib05}, in agreement with our analysis here which concluded 
that the excited $t1$ triplet is a realistic candidate for acting as a doorway 
state.  

\begin{figure}[t]
\centering{\includegraphics[width=7.30cm]{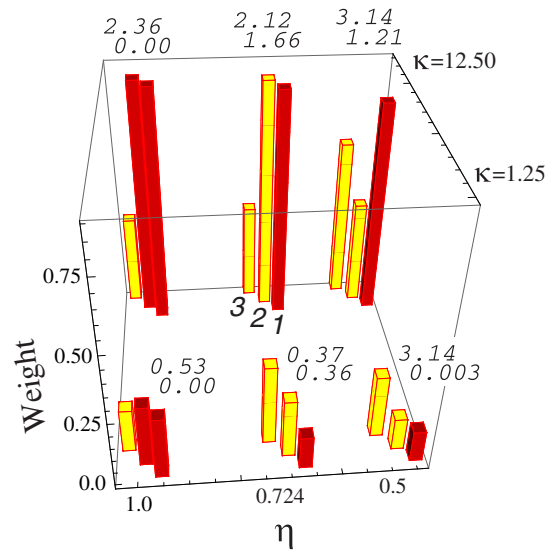}}
\caption{(color online)
The bar-chart for the relevant quasiparticles of the $N=2 \rightarrow N=3$ 
transition in a quantum dot as a function of the strength of the interelectron 
repulsion (through the dielectric constant $\kappa$) and the anisotropy $\eta$. 
The heights of the bars indicate the weight ${\cal W}$ [Eq.\ (\ref{wqp})], with
a dark shade (red) denoting $\Delta \theta_{\text{QP}}=\pi$ and a gray shade 
(yellow) denoting $\Delta \theta_{\text{QP}}=0$. 
In all instances the initial state is 
the singlet [$(S=0,S_z=0)$] ground state of the $N=2$ dot.
The bars are arranged in groups of three around each ($\kappa$,$\eta$) point; 
the rightmost bar corresponds to the final $N=3$ ground state (index 1), while 
the middle (index 2) and leftmost (index 3) ones correspond to the final $N=3$ 
two lowest excited states with a non-zero weight. All three final states happen 
to have a $(S=1/2,S_z=1/2)$ magnetic structure.
The numbers above the bars denote the energy difference (in meV) between the 
$N=3$ ground-state and the two lowest excited $(S=1/2,S_z=1/2)$ states.
The dot parameters are as in Fig.\ \ref{figbars12}.
}
\label{figbars23}
\end{figure}

For the $N =2 \rightarrow N=3$ transition, the initial many-body 
state is assumed to be the ground state of a QD with two electrons, which is 
always a singlet [$(S=0,S_z=0)$]. Again the final state cannot be 
restricted to the ground state [with ($S=1/2,S_z=1/2$)] of the $N=3$ system 
\cite{mikh02li07}, since excited states may act as doorway states. 
Because the transitions to a final $(S=3/2,S_z=3/2)$ or 
$(S=3/2,S_z=1/2)$ state are forbidden (the corresponding ${\cal W}=0$ due to
spin blockade), we are led to consider three final states, i.e., the 
ground-state with $(S=1/2,S_z=1/2$) and the two-lowest excited states also with 
$(S=1/2,S_z=1/2)$ \cite{mikh02li07}. In Fig.\ 
\ref{figbars23}, these final states are denoted by the indices 1, 2,
and 3, respectively; our results apply unaltered for final states 
with an $S_z=-1/2$ spin projection. 

In Fig.\ \ref{figbars23}, we retain the same conventions as in Fig.\ 
\ref{figbars12} concerning the height and shadings (colors online) of the
bars. Unlike Fig.\ \ref{figbars12}, the final two-lowest excited states in Fig.\
\ref{figbars23} are not degenerate, and thus in the latter case we list a pair 
of energy gaps (in meV) with respect to the final ground state. 
In the circular case ($\eta=1$), there are two degenerate final ground states
(with total angular momentum $L=1$ and $L=-1$) \cite{mikh02li07}, and the 
smallest meaningful excitation energy gap needs to be taken between the states 
with indices 1 and 3 (or 2 and 3).

Inspection of Fig.\ \ref{figbars23} reveals that a transition to the $N=3$ 
ground state (marked 1) will always have $\Delta \theta_{\text{QP}}=\pi$ 
[corresponding to a dark shade (online red)], 
and thus it will exhibit no lapse in the
transmission phase $\theta_{\text{PL}}$ [see Eq.\ (\ref{the})], 
in contrast to the experimental result. 
Transmission through a doorway excited state may become possible for smaller
$\kappa$ (stronger Coulomb repulsion) and smaller $\eta$ (stronger anisotropy),
since as a result (1) the energy gap between the first excited state (index 2) 
and the ground state diminishes (observe the practically vanishing gap of
0.003 meV for $\eta=0.5$ and $\kappa=1.25$), and (2) the weight ${\cal W}$ of 
this index-2 final state remains larger than that of the ground state
[see the cases with ($\kappa=1.25$, $\eta=0.724$) and ($\kappa=1.25$, 
$\eta=0.5$)]. Transmission through such an index-2 doorway state will lead to a 
phase lapse ($\theta_{\text{PL}}=-\pi$), since in this case 
$\Delta \theta_{\text{QP}}=0$ 
[gray shade (online yellow)]. A phase lapse in the $N=2 \rightarrow N=3$ 
transition has indeed been observed \cite{heib05}. 
As for the $N=1 \rightarrow N=2$ transition, this 
observation of a phase lapse and our EXD analysis of the $N=2 \rightarrow N=3$ 
transition suggest that the dots in the experiments were strongly deformed 
and exhibited rather strong ineterelectron correlations.

{\it Conclusions.\/}
In summary, focusing on the mesoscopic regime of electron interferometry, 
and using the Bardeen weak-coupling theory in conjunction 
with exact diagonalization of the 
many-body quantum dot Hamiltonian, we have shown for the first two transitions, 
(a) $N=1 \rightarrow N=2$ and (b) $N=2 \rightarrow N=3$, nonuniversal behavior 
of the transmission phases with no phase lapse for (a) and a phase lapse of 
$\pi$ for (b), in agreement with the experiment \cite{heib05}. These results 
were obtained for a range of dot parameters characterized by shape anisotropy 
and strong $e-e$ repulsion, with both favoring electron localization and 
formation of Wigner molecules \cite{rop07,note2,note3,note4}. 
Additionally, our analysis of the 
quasiparticle wavefunction [Eq.\ (\ref{wfqp})] highlights the dependence of the 
phase-lapse behavior on the spin configurations of the initial and final states,
and the importance of excited doorway states as favored transmission channels. 
Electron interferometric measurements on dots with characterized 
shapes \cite{note5}, as well as extension of our analysis to a larger number 
of electrons and the transition to the universal regime, including
stronger lead-dot coupling and possibly explicit incorporation of 
lead states, remain future challenges.

Work supported by the US D.O.E. (Grant No. FG05-86ER45234).

\vfill

\end{document}